\documentclass[twoside]{dis07}
\usepackage[latin1]{inputenc}
\usepackage[dvips]{graphicx,epsfig,color}
\usepackage{wrapfig,rotating}
\usepackage{amssymb,amsmath,array}
\def\T{{\cal T}}
\def\N{{\cal N}}
\def\g{\gamma}
\def\f{\frac}

\begin{document}
\title{Non-forward Balitsky-Kovchegov equation and Vector Mesons}

\author{Robi Peschanski$^1$, Cyrille Marquet$^2$ and Gregory Soyez$^3$
%
%
\vspace{.3cm}\\
%
1- Service de Physique Th\'eorique - CEA/Saclay,
91191 Gif-sur-Yvette Cedex, FRANCE
%
\vspace{.1cm}\\
2- RIKEN BNL Research Center, Brookhaven National Laboratory, 
Upton, NY 11973, USA
\vspace{.1cm}\\
3- Physics Department, Brookhaven National Laboratory,
Upton, NY 11973, USA
\vspace{.1cm}\\
}

\maketitle

\begin{abstract}
  Considering the Balitsky-Kovchegov QCD evolution equation in full
  momentum space, we derive the travelling wave solutions expressing
  the nonlinear saturation constraints on the dipole scattering
  amplitude at non-zero momentum transfer. A phenomenological
  application to elastic vector meson production shows the
  compatibility of data with the QCD prediction:  an \emph{enhanced}
  saturation scale at intermediate momentum transfer.
\end{abstract}

\section{Motivation}

The saturation of parton densities at high energy has been mainly
studied for the forward dipole-target scattering amplitude
${\T}(r,q=0,Y),$ where $r,q,Y$ are, respectively, the dipole size, the
momentum transfer and the total rapidity of the process. For instance,
the corresponding QCD Balitsky-Kovchegov (BK) equation \cite{bk} has
been shown to provide a theoretical insight on the ``geometric
scaling'' properties \cite{gs} of the related $\g^*$-proton
cross-sections. Indeed, it can be related to the existence of a
scaling for ${\T}(r,q=0,Y)\sim {\T}(r^2 Q^2(Y))$ where the saturation
scale is $Q^2(Y)\sim \exp{c Y}$ and the constant $c$ can be
interpreted as the critical speed of ``travelling wave'' solutions of
the nonlinear BK equation \cite{tw}. Our theoretical and
phenomenological subjects are the extension of these properties to the
non-forward amplitude ${\T}(r,q\ne 0,Y),$ which is  
phenomenologically relevant
for the elastic production of vector mesons in deep inelastic
scattering.

\section{BK equation in full momentum space} 
\begin{wrapfigure}{r}{0.5\columnwidth}
\vspace*{-0.1cm}
\centerline{\includegraphics[width=0.43\columnwidth]{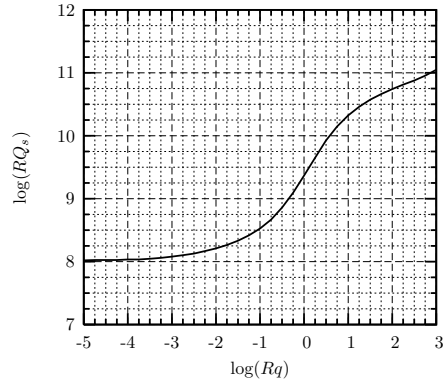}}
\caption{$q^2-$dependent saturation scale}
\label{1}
\end{wrapfigure}
In order to study the properties of ${\T}(r,q\ne 0,Y),$ one has first
to deal with both conceptual and technical difficulties. It is known
that the BK formalism has been originally derived in impact parameter
$b$ but then its validity especially at large $b$ is questionable, 
since it leads to non physical power-law tails.
Hence we start with the formulation of the BK equation in momentum
$q$, which is more \emph{local} but has a non-trivial nonlinear form \cite{bkfull}. In fact,
despite this problem, the general method of travelling wave solutions
can be extended in the non-forward domain \cite{us}. It consists in 3
steps: first, one solves the equation restricted to its linear part
which is related to the non-forward Balitsky Fadin Kuraev Lipatov
(BFKL) equation \cite{bfkl} for the dipole-dipole amplitude {\em via}
factorisation and whose solution takes the form of a linear
superposition of waves. Second, one finds that the nonlinearities act
by selecting the travelling wave with {\it critical} speed $c,$ in a
way which, interestingly, is independent of the specific structure of
the nonlinear damping terms. Third, one obtains after enough rapidity
evolution, a solution which appears independent from initial
conditions ($\T_0 \sim r^{2 \g_0}$) , provided these are sharper than
the critical travelling wave front profile $\T \sim r^{2 \g_c},$ with
$\g_0 > \g_c.$ Interestingly enough, QCD color transparency satisfies 
this criterium.
\begin{figure}[ht]
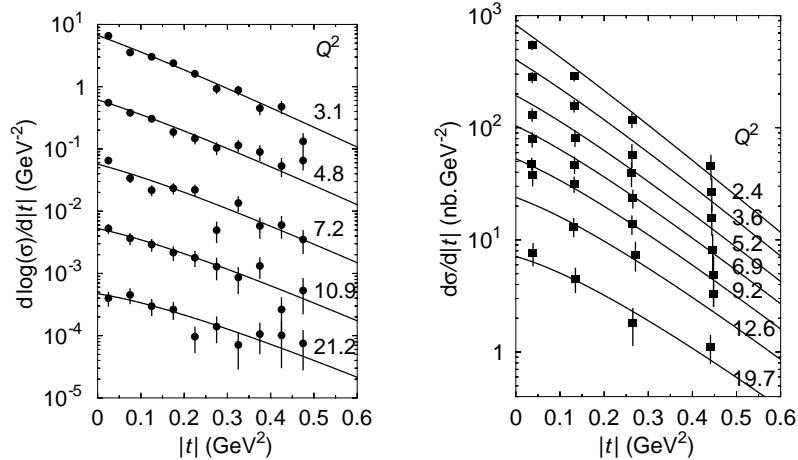

\centerline{\includegraphics[width=0.35\columnwidth]{peschanski_robi.fig2.ps}
\hspace*{0.5cm}
\includegraphics[width=0.35\columnwidth]{peschanski_robi.fig3.ps}}
\caption{$\rho$ (H1) and $\phi$ (ZEUS) differential 
cross-sections at $W=75\ {\rm GeV}$}\label{2}
\end{figure}
Applying these general results on the non-forward case one finds the
following QCD predictions, depending on the relative magnitude of
three scales involved in the process, namely $q,$ $k_T^{-1}$ (the
target size) and $k_P^{-1}\equiv r$ (the projectile \emph{i.e.} dipole
size).
\begin{itemize}
\item Near-Forward region $q \ll k_{T} \ll k_P:\ Q_s^2(Y) \sim
  k_{T}^2\exp{c Y}$
\item Intermediate transfer region $k_{T} \ll q \ll k_P:\ Q_s^2(Y)
  \sim q^2\exp{c Y} $
\item High transfer region $q \ll k_{T} \ll k_P: $ No saturation.
\end{itemize}
Our main prediction is thus the validity of the forward travelling
wave solution extended in the non-forward intermediate-transfer domain
but with an {\it enhanced} saturation scale by the ratio
$q^2/k_{T}^2,$ where $k_{T}$ is a typically small, nonperturbative
scale. Hence we are led to predict \emph{geometric scaling} properties
with a purely perturbative initial saturation scale given by the
transverse momentum. This saturation scale {\it enhancement}
prediction is confirmed by numerical simulations of the BK
solutions as shown in Fig.\ref{1}.

\section{QCD Saturation Model for Exclusive VM production}

The differential cross-section for exclusive vector meson (VM)
production at HERA, see Fig.{2}, can be theoretically 
obtained from the non-forward
 dipole-proton
amplitude and from $\Phi_{T,L}^{\g^*V},$ the overlap functions between
the (longitudinal and  transverse) virtual photon and vector
meson wave-functions \cite{pheno}. For completion, we used two
different VM wave-functions of the literature, without noticeable
difference in our conclusions. One writes
\begin{eqnarray}
\f{d\sigma^{\g^*p\rightarrow Vp}_{T,L}}{d{q^2}}&=&\f1{16\pi}
\left|\int d^2r \int_0^1 dz\ {\Phi_{T,L}^{\g^*V}(z,{r};Q^2,M_V^2)}\
e^{-iz{{q}}\cdot{r}}\ \T ({r},{q},Y)\right|^2\ ,\nonumber
\end{eqnarray}

Following theoretical prescriptions, we consider a forward
dipole-proton amplitude $\N_{IIM}$ satisfactorily describing the total
DIS cross-sections in a saturation model \cite{iim}. We just make the
saturation scale varying with $q^2,$ following the trend shown in
Fig.\ref{1} and starting from the forward model one $Q_s^2(Y),$ one
writes
\begin{eqnarray}
{{ {T}({r},{{q}};Y)= 2\pi R_p^2\ e^{-B {{ q^2}}}
\N_{IIM}({r}^2\ Q_s^2(Y,{{q}})) }}\ ;\ Q_s^2(q,Y) = Q_s^2(Y)\ (1+c\ {q^2})\ .
\nonumber
\end{eqnarray}
\begin{wraptable}{l}{0.5\columnwidth}
\centerline{\begin{tabular}{|c|c|c|}
\hline
Cross-sections  & $q^2$-Sat. &  fixed-Sat. \\\hline
$\rho,\ \sigma_{\text{el}_{}}$  & 1.156 & 1.732 \\\hline
$\rho,\ \frac{^{}d\sigma}{dt_{}}$  & 1.382 & 1.489 \\\hline  
$\phi,\ \sigma_{\text{el}_{}}$  & 1.322 & 2.247 \\\hline
$\phi,\ \frac{^{}d\sigma}{dt_{}}$  & 1.076 & 0.931 \\\hline\hline
Total  & 1.212 & 1.480 \\\hline
\hline
\end{tabular}}
\caption{Comparison of the $\chi^2/{\rm points}$}
\label{tab}
\end{wraptable}
\noindent The factor $2\pi R_p^2\ e^{-B {{ q^2}}}$ comes from the
non-perturbative proton form factor. For clarity of the analysis, we
considered only $B$ and $c$ as free parameters of the non-forward
parametrisations, the others being independently fixed by the forward
analysis.

In Table \ref{tab}, one displays the $\chi^2/{\rm point}$ obtained by
a fit of $\rho$ (47 data points) and $\phi$ (34 points) total elastic
production cross-sections and of $\rho$ (50 data points) and $\phi$
(70 points) differential cross-sections. The Table compares the
saturation fits for fixed and $q^2$-dependent scales, with a favour
for the enhanced-scale model in the total. The model gives a
comparable fit with a more conventional non-saturation model using a
$Q^2$-dependent slope $B\propto M_V^2+Q^2.$ Some of our results for
the cross-sections are displayed in the figures.  In Fig.\ref{2}, one
shows the results of the fit for $\rho$-production (H1) and
$\phi$-production (ZEUS) differential cross-sections for a total
$\g^*-p$ energy $W=\ 75 {\rm GeV}$ and different $Q^2$ values.
\begin{figure}[ht]
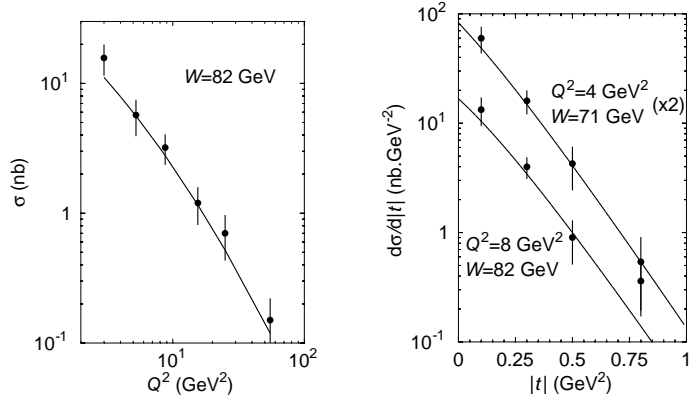

\centerline{\includegraphics[width=0.3\columnwidth]{peschanski_robi.fig4.ps}
\hspace*{0.5cm}
\includegraphics[width=0.3\columnwidth]{peschanski_robi.fig5.ps}}
\caption{Predictions for the DVCS measurements. Left plot:
  cross-section, right plot: differential cross-section.}\label{4}
\end{figure}
Let us finally present our predictions for the DVCS cross-section,
which is obtained without any free parameter from our analysis. In
Fig.\ref{4}, they are compared with the available data and the
agreement is good in the simple chosen parametrisation.

\section{Conclusions}
Let us summarize our new results

$\bullet$ \emph {Saturation at non-zero transfer:}
The Balitsky-Kovchegov QCD evolution  equation involving  
full momentum transfer predicts (besides the known $q=0$ case) saturation in the 
\emph{intermediate} transfer range, namely for $Q_0<q<Q,$ where $Q_0$ 
(resp. $Q$) is the target (resp. projectile) typical scale.

$\bullet$  \emph {Characterisation of the universality class:} 
The universality class of the corresponding travelling-wave solutions  is governed by a purely perturbative
saturation scale $Q_s(Y)\equiv q^2 \Omega(Y),$ where $\Omega(Y) 
\sim e^{cY}$ is the same 
rapidity evolution factor as in the forward case. Consequently the
\emph {intermediate transfer} saturation scale gets enhanced by a factor 
$q^2/Q_0^2.$ 

$\bullet$  \emph{Phenomenology of Vector mesons:}
The QCD predictions are applied in the experimentally accessible 
\emph {intermediate transfer} range of vector meson production. 
The model uses an interpolation between the forward and non-forward saturation
 scale together with a parameter-frozen forward saturation model. It fits better the data on 
$\rho$ (H1) and $\phi$ (ZEUS)  
cross-sections than for a non-enhanced saturation. 

$\bullet$ \emph {Prospects:} The next phenomenological prospect is to
add charm to the discussion, both with the modification of the forward
case by including the charm contribution \cite{F2hq} and by also
considering the production of $\Psi$ mesons. On a theoretical ground,
it would be interesting to go beyond
the mean-field approximation of the BK equation.\\
 
\begin{footnotesize}




\begin{thebibliography}{99}
\bibitem{url} Slides: \\ 
\verb$http://indico.cern.ch/contributionDisplay.py?contribId=75&sessionId=7&confId=9499$
\bibitem{bk}
I. Balitsky,
{\it Nucl. Phys.} {\bf B463} (1996) 99;
{\it Phys. Lett.} {\bf B518} (2001) 235;\\
Yu.V. Kovchegov,
{\it Phys. Rev.} {\bf D60} (1999) 034008;
{\it Phys. Rev.} {\bf D61} (2000) 074018.

\bibitem{gs}
A.M. Stasto, K. Golec-Biernat and J. Kwiecinski,
{\it Phys. Rev. Lett.} {\bf 86} (2001) 596;\\
C. Marquet and L. Schoeffel,
{\it Phys. Lett.} {\bf B639} (2006) 471. 

\bibitem{tw}
S. Munier and R. Peschanski,
{\it Phys. Rev. Lett.} {\bf 91} (2003) 232001;
{\it Phys. Rev.} {\bf D69} (2004) 034008;
{\bf D70} (2004) 077503.

\bibitem{bkfull}
C. Marquet and G. Soyez,
{\it Nucl. Phys.} {\bf A760} (2005) 208.

\bibitem{us}
C. Marquet, R. Peschanski and G. Soyez,
{\it Nucl. Phys.} {\bf A756} (2005) 399.

\bibitem{bfkl}
  L.~N.~Lipatov,
  Sov.\ Phys.\ JETP {\bf 63}, 904 (1986)
  [Zh.\ Eksp.\ Teor.\ Fiz.\  {\bf 90}, 1536 (1986)].

\bibitem{pheno}
C. Marquet, R. Peschanski and G. Soyez, 
{\it Exclusive vector meson production at HERA from QCD with saturation},
hep-ph/0702171.

\bibitem{iim}
E. Iancu, K. Itakura and S. Munier,
{\it Phys. Lett.} {\bf B590} (2004) 199.
  
\bibitem{F2hq}
K. Golec-Biernat and S. Sapeta,
{\it Phys. Rev.} {\bf D74} (2006) 054032;\\
H. Kowalski, L. Motyka and G. Watt,
{\it Phys. Rev.} {\bf D74} (2006) 074016;\\
G.~Soyez, {\em Saturation QCD predictions with heavy quarks at HERA},
arXiv.0705.3672.

\end{thebibliography}
%

\end{footnotesize}

\end{document}